\documentclass[11pt]{aastex}
\usepackage{emulateapj5}
\usepackage{apjfonts}
\usepackage{epsfig}

\newcommand{\BH}{\ensuremath{\mathrm{BH}}}
\newcommand{\MBH}{\ensuremath{M_\mathrm{BH}}}

\newcommand{\RBH}{\ensuremath{R_\mathrm{BH}}}

\newcommand{\Msph}{\ensuremath{M_\mathrm{bul}}}
\newcommand{\Mdyn}{\ensuremath{M_\mathrm{dyn}}}
\newcommand{\Lsph}{\ensuremath{L_\mathrm{bul}}}

\newcommand{\LBsph}{\ensuremath{L_\mathrm{B,bul}}}

\newcommand{\LKsph}{\ensuremath{L_\mathrm{K,bul}}}
\newcommand{\LNIRsph}{\ensuremath{L_\mathrm{NIR,bul}}}

\newcommand{\sigstar}{\ensuremath{\sigma_e}}

\newcommand{\1}{\ensuremath{^{-1}}}

\newcommand{\KM}{\ensuremath{\mathrm{~km}}}

\newcommand{\MPC}{\ensuremath{\mathrm{~Mpc}}}

\newcommand{\SEC}{\ensuremath{\mathrm{~s}}}

\newcommand{\mlr}{\ensuremath{\Upsilon}}

\newcommand{\eg}{e.g.,}
\newcommand{\ie}{i.e.,}

\shorttitle{Black Hole Masses and Near-Infrared Bulges}
\shortauthors{Marconi \& Hunt}

\slugcomment{Astrophysical Journal Letters, in press }

\begin{document}

\title{The Relation between Black Hole Mass, Bulge Mass, and
Near-Infrared Luminosity}

\author{
        Alessandro~Marconi\altaffilmark{1} and
	Leslie~K.~Hunt\altaffilmark{2}
}
\altaffiltext{1}{INAF- Osservatorio Astrofisico di Arcetri,
L.go Fermi 5, I-50125 Firenze, Italy; marconi@arcetri.astro.it.}
\altaffiltext{2}{Istituto di Radioastronomia-Sez.\ Firenze/CNR, L.go Fermi 5, I-50125 Firenze, Italy; hunt@arcetri.astro.it.}

\begin{abstract}
We present new accurate near-infrared (NIR) spheroid (bulge)
structural parameters obtained by two-dimensional image
analysis for all galaxies with a direct black
hole (\BH) mass determination.  As expected, NIR bulge luminosities \Lsph\ and
\BH\ masses are tightly correlated,  and if we consider only those galaxies
with secure \BH\ mass measurement and accurate \Lsph\ (27 objects),
the spread of \MBH-\Lsph\ is similar to \MBH-\sigstar, where \sigstar\ is
the effective stellar velocity dispersion.  We find an 
intrinsic $rms$ scatter of $\simeq 0.3$\,dex in log\,\MBH.
By combining the bulge effective radii $R_e$ measured in our analysis with
\sigstar, we find a tight linear correlation ($rms \simeq 0.25$ dex) between \MBH\
and the virial bulge mass ($\propto R_e \sigstar^2$), with
$\langle \MBH/\Msph\rangle \sim 0.002$.  A partial correlation
analysis shows that \MBH\ depends on both \sigstar\ and 
$R_e$, and that both variables are necessary to drive
the correlations between \MBH\ and other bulge properties.
\end{abstract}


\section{\label{sec:intro}Introduction}

Central massive black holes (\BH s) are now thought to reside in virtually 
all galaxies with a hot spheroidal stellar component (hereafter bulge). 
Such \BH s seem to be a relic of past quasar activity (\eg\
\citealt{soltan,ms02,yu02,ar02}) and related
to host galaxy properties, with the implication that \BH\
and galaxy formation processes are closely linked.  
Previous work has shown that
\BH\ mass \MBH\ is correlated with both blue luminosity \LBsph\ and bulge mass \Msph, although with
considerable intrinsic scatter ($rms\sim 0.5$ in log\MBH; \citealt{kr95}).
However, \MBH\ and the bulge effective stellar velocity
dispersion \sigstar\ 
correlate more tightly ($rms\sim 0.3$)
than \MBH-\LBsph\ \citep{fm00,gebhardt00}. 
The smaller scatter of the \MBH-\sigstar\ correlation 
suggests that the bulge
dynamics (or mass), rather than luminosity, is the agent of the correlation.
But the smaller spread relative to \MBH-\Lsph\ appears to be an 
artefact of the manipulations necessary to derive \Lsph.
Indeed, recent work has shown that when
bulge parameters are measured with more accuracy 
[e.g. profile fitting
rather than average correction for disk light \citep{simien}],
the resulting scatter is comparable to that
of \MBH-\sigstar\ \citep{mclure02,erwin}.
The correlation between \MBH\ and bulge light concentration
also has a comparably low scatter \citep{graham01}.
Nevertheless, there are strong indications that \LBsph\ of the brightest elliptical
galaxies, for which decomposition issues are unimportant,
deviate significantly from the \MBH-\Lsph\ relation \citep{ferrarese02}. 
Hence, longer wavelengths may also be necessary to better define the
intrinsic scatter in \MBH-\Lsph\ compared to that of \MBH-\sigstar.

In this paper, we reexamine the \MBH-\Lsph\ correlation
by accurately measuring the bulge
luminosity in the near-infrared (NIR) for all galaxies with a well-determined
\MBH.
All previous studies have used optical light (B or R) to test
the \MBH-\Lsph\ relation, but
NIR light provides a clear advantage over the optical: it is a 
better tracer of stellar mass and less subject to the effects of extinction.
If the physical correlation is between the \BH\ mass and bulge mass,
the NIR correlations \MBH-\Lsph\ should be tighter than those in the optical,
because of the smaller variation of $M/L$ ratio \mlr\ with mass
(\eg\ \citealt{gavazzi}).
Moreover, we use a two-dimensional (2D) bulge/disk decomposition to determine
bulge parameters, an improvement on earlier work which applied 1D
fits only.
Here we construct the largest possible sample, by considering all galaxies 
which have been used for the \MBH-\sigstar\ and \MBH-\LBsph\ correlations.
In \S\ref{sec:sample} we present the sample of galaxies with direct
dynamical \BH\ mass measurements, and in \S\ref{sec:analysis} 
describe the images and the 2D bulge/disk decomposition applied to them.
Finally,
in \S\ref{sec:results} we discuss the results of the analysis.

\begin{figure*}[t]
\centering
\epsfig{file=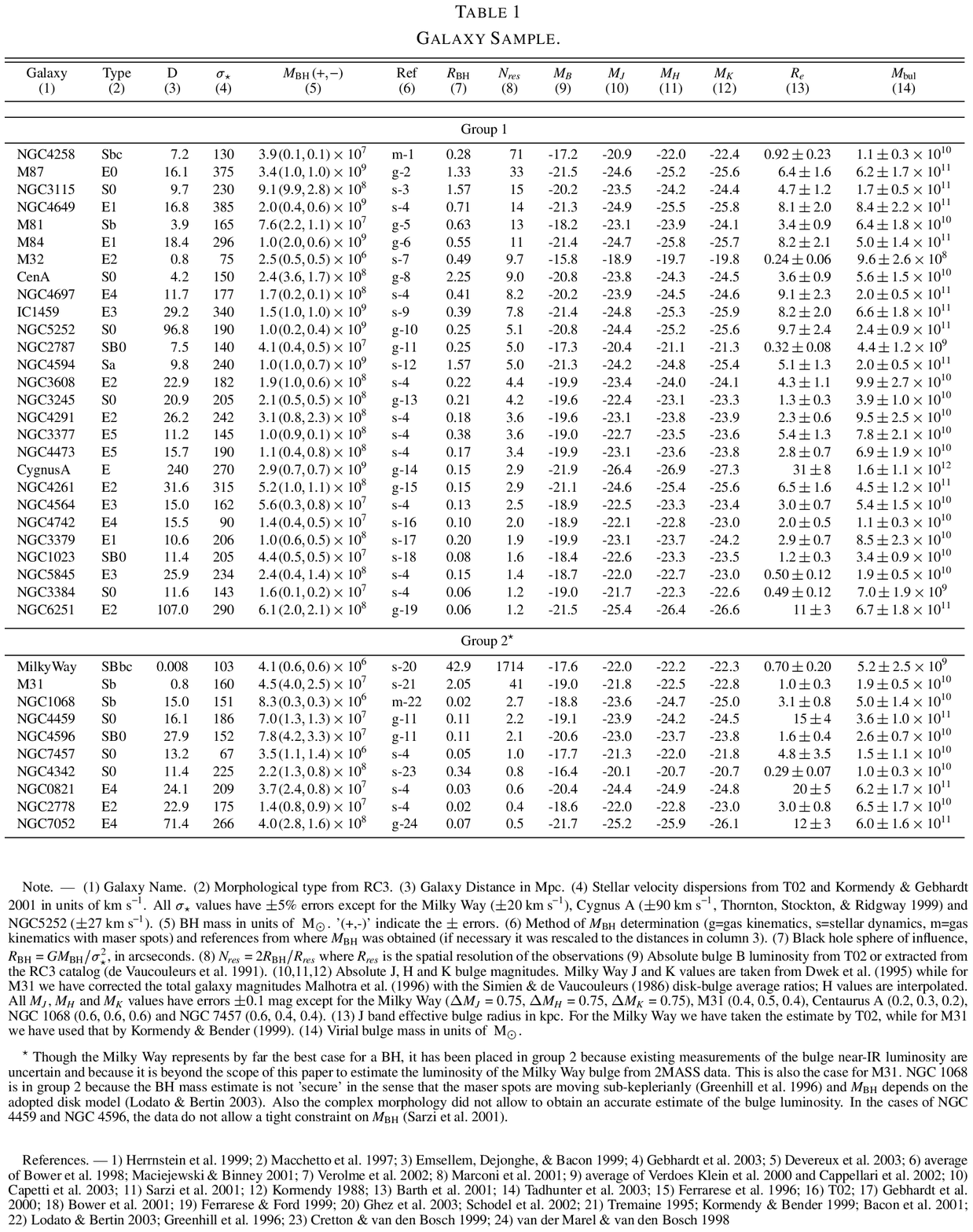}
\end{figure*}

\section{\label{sec:sample} The Sample}

To date, there are 37 galaxies with direct gas kinematical or stellar dynamical
determination of the central \BH\ mass. These galaxies have been compiled and
made into a uniform sample (\eg\ for distances) by a number of authors (\eg\
\citealt{mf02,tremaine02}; hereafter MF02 and T02, respectively).
We adopt the data from the recent paper by
T02 with some modifications and additions.  The data in Columns
1-5 and 9 of Table 1 are from the compilation by
T02 and the reader can refer to that paper for more details.  
Differently from
T02, when galaxy distances from surface brightness
fluctuations (\citealt{tonry}) are not available, we use recession
velocities corrected for Virgocentric infall from
the LEDA database (\url{http://leda.univ-lyon1.fr/})
with $H_0=70\KM\SEC\1\MPC\1$.
In a few cases, we also consider \BH\ mass estimates from different
papers than those used by T02; thus, in Col.\ 6,
we indicate the appropriate references.
With respect to the 31 galaxies considered by T02
we add: Cygnus A \citep{tadhunter03}, M81 \citep{devereux03},
M84 \citep{bower98}, NGC 4594 \citep{k88}, Centaurus A
\citep{marconi01} and NGC 5252 \citep{capetti03}.

Following MF02, we divide the galaxies into two groups.
In the first group, we place all the galaxies which have a secure \BH\ mass
measurement and an accurate determination of the bulge NIR luminosity. We
consider `secure' those \BH\ masses for which the black hole sphere of
influence, $\RBH=G\MBH/\sigstar^2$ (column 7 of Table 1), has
been clearly resolved, \ie\ $N_{res}=2\RBH/R_{res}>1$,
where $R_{res}$ is the spatial resolution
of the observations.
Additional reasons for placing galaxies in Group 2 are given in Table 1.
\begin{figure*}[t]
\centerline{
\epsfig{file=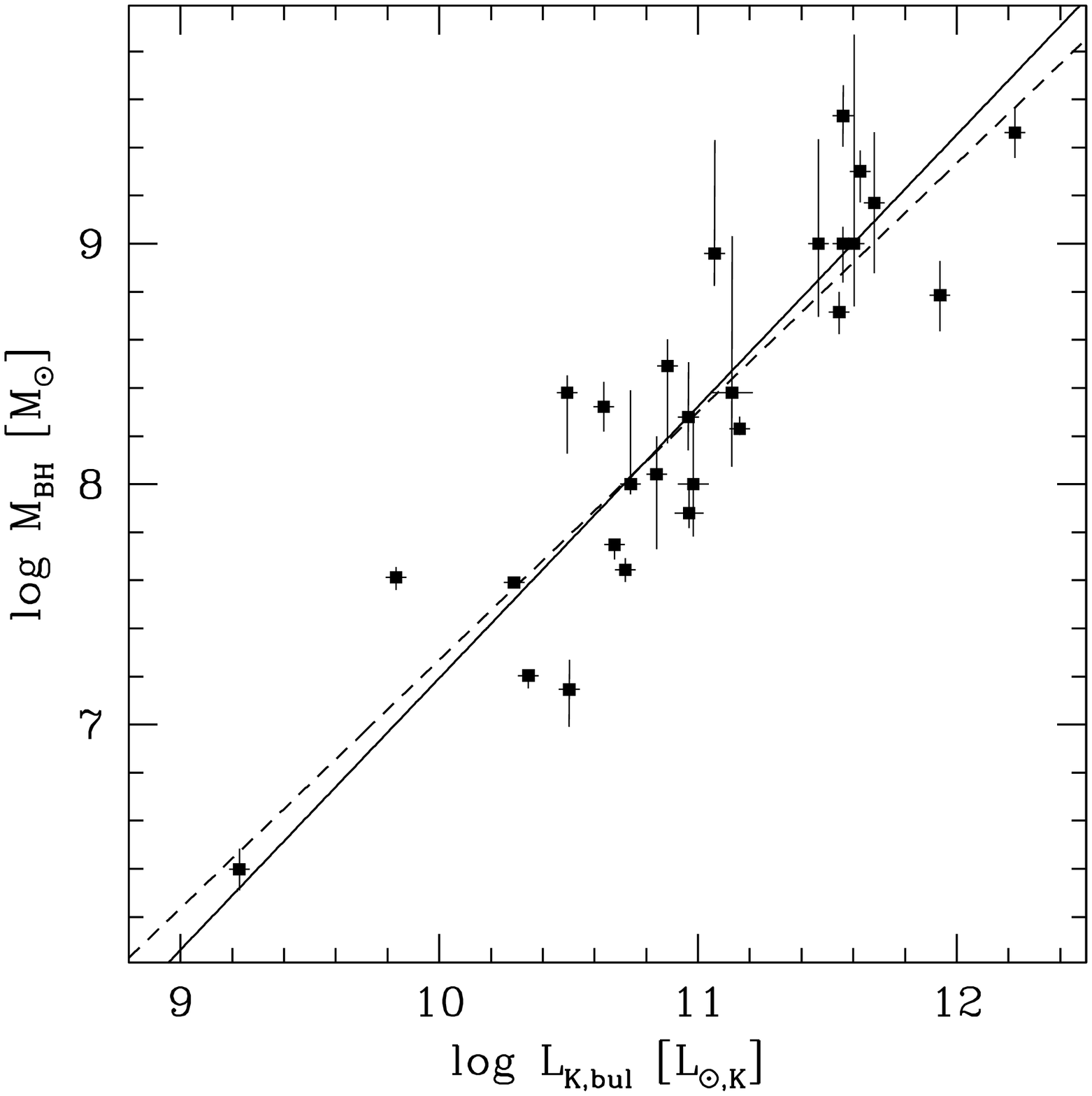, width=0.33\linewidth}
\epsfig{file=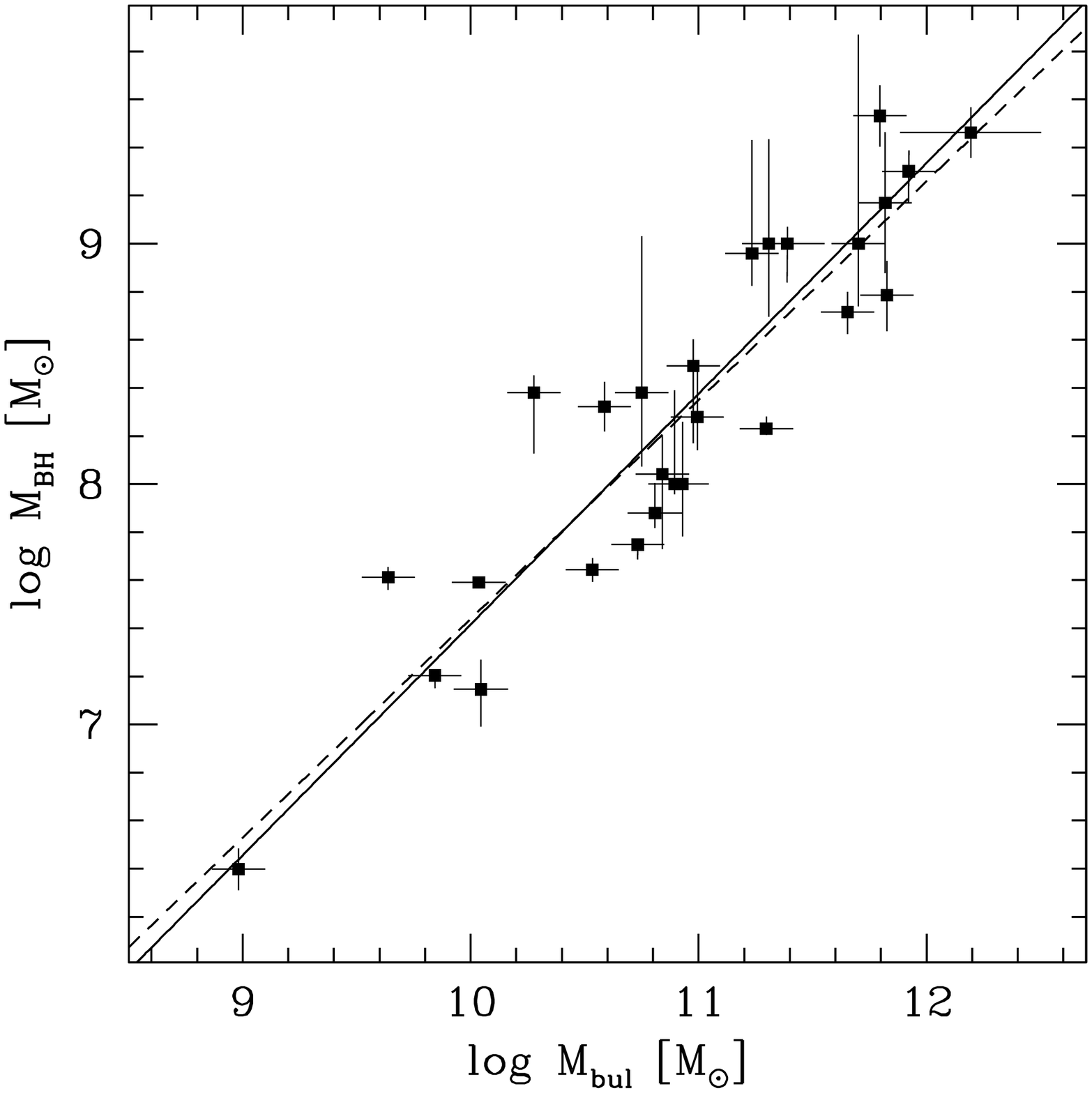, width=0.33\linewidth}
\epsfig{file=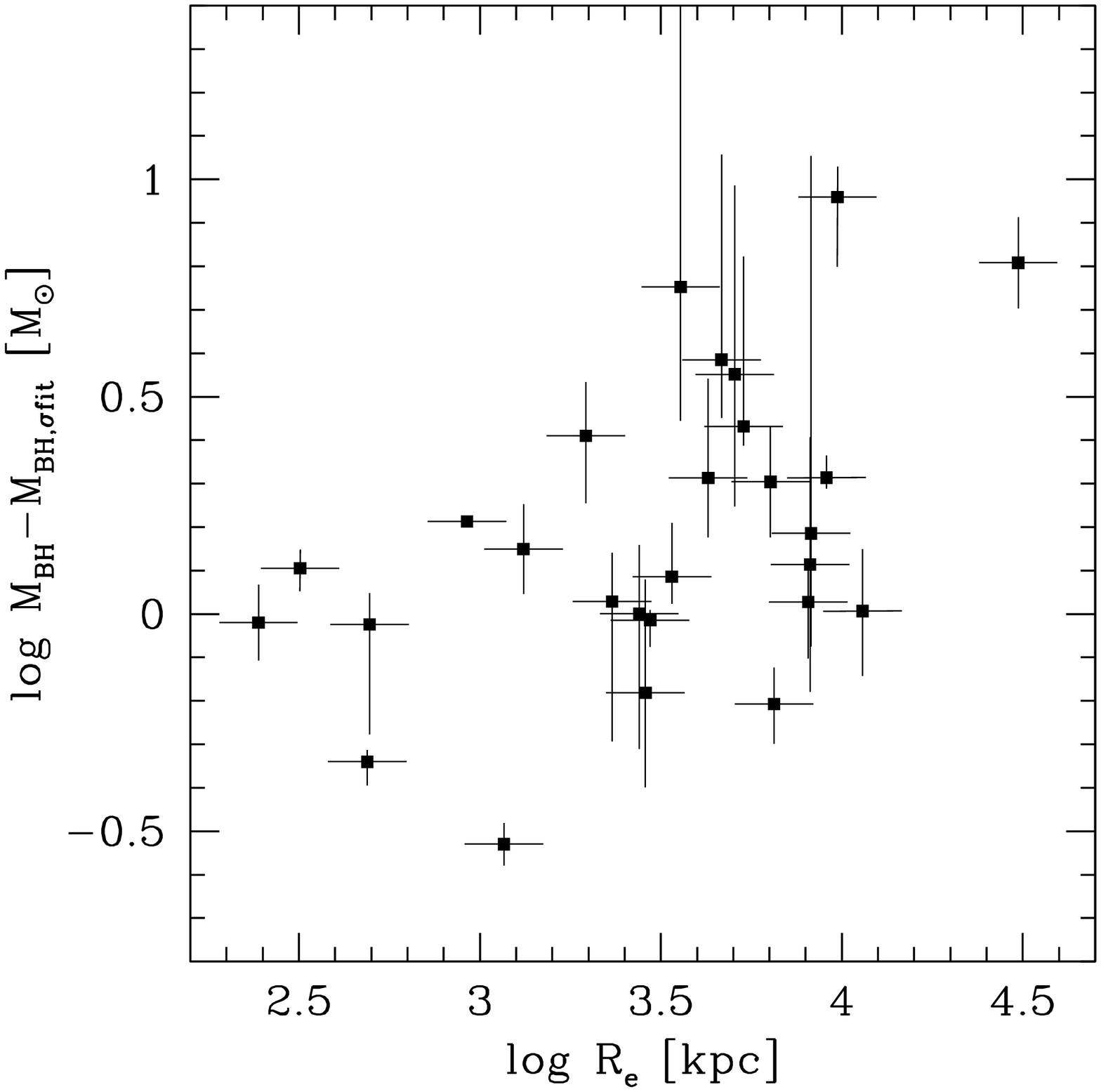, width=0.33\linewidth}
}
\caption{\label{fig:corr} Left (a): \MBH\ vs \LKsph\ for the galaxies of Group 1.
The solid lines are obtained with the bisector linear regression algorithm by
\cite{ab}, while
the dashed lines are ordinary least square fits.
Center (b): \MBH\ vs \Msph\ with the same notation as in the previous
panel. Right (c): residuals of \MBH-\sigstar\ vs $R_e$, in which we use the \MBH-\sigstar\
regression by T02.} 
\end{figure*}

\section{\label{sec:analysis} Image Analysis}

We have constructed a homogeneous set of NIR images of the galaxies presented
in Table 1 (except for the Milky Way and M31)
by retrieving J, H, and K atlas images
from the 2-Micron All Sky Survey (2MASS; \url{http://www.ipac.caltech.edu/2mass}). 
When a single atlas image contained only a
portion of the galaxy, we also retrieved adjacent tiles and mosaicked the
images after subtracting the sky background
and rescaling for the different zero points.
The 2MASS images are photometrically
calibrated with a typical accuracy of a few percent.  More details can be
found in Hunt \& Marconi (2003; hereafter Paper~II).

We performed a 2D bulge/disk decomposition of the images using the
program GALFIT \citep{peng} which is made publicly available by the authors.
This code allows the fitting of several components with different functional
shapes (e.g., generalized exponential (Sersic) and simple exponential laws);
the best fit parameters are determined by
minimizing $\chi^2$.  More details on GALFIT can be found in \cite{peng}.
We fit separately the J, H and K images. Each fit was started by fitting a
single Sersic component and constant background. When necessary (\eg\ for
spiral galaxies), an additional component (usually an exponential disk) was
added.  In many cases these initial fits left large residuals and we thus
increased the number of components (see also \citealt{peng}).
The fits are described in detail
in Paper~II.  In Table 1 we present
the J, H and K bulge magnitudes, effective bulge radii $R_e$ in the J band,
and their uncertainties. The J, H and K magnitudes were corrected for Galactic
extinction using the data by \cite{galabs}.  We used the J band to determine
$R_e$ because the images tend to be flatter, and thus the background is better
determined.

\section{\label{sec:results} Results and Discussion}

In Fig.\ \ref{fig:corr} we plot, from left to right, \MBH\ vs \LKsph,
\MBH\ vs \Msph, and the residuals of \MBH-\sigstar\ vs $R_e$ (based on
the fit from T02).
Only Group 1 galaxies are shown.  \Msph\ is the virial bulge mass
given by $k\,R_e\,\sigstar^2/G$;
if bulges behave as isothermal spheres, $k=8/3$. 
However, comparing our virial estimates \Msph\ with those \Mdyn, obtained
from dynamical modeling \citep{magorrian,gebhardt03}, 
shows that \Msph\ and \Mdyn\ are
well correlated ($r=0.88$); setting $k=3$ (rather
than $8/3$) gives an average ratio of unity.
Therefore, we have used $k=3$ in the above formula. 
Considering the uncertainties of both mass estimates, the scatter
of the ratio \Msph/\Mdyn\ is 0.21 dex.
We fit the data with the bisector linear regression 
from \cite{ab} which allows for uncertainties
on both variables and intrinsic dispersion.  The 
FITEXY routine \citep{numrec} used by T02
gives consistent results (see Fig.\ \ref{fig:corr}). Fit results of \MBH\ vs
galaxy properties for Group 1 and the combined samples are summarized in Table 2.  The intrinsic
dispersion of the residuals ($rms$) has been estimated with a maximum
likelihood method assuming normally-distributed values.
Inspection of Fig.\ \ref{fig:corr} and Table 2 show that \LKsph\
and \Msph\ correlate well with the \BH\ mass.
The correlation between \MBH\
and \Msph\ is equivalent to that between the radius of the \BH\ sphere of
influence \RBH\ ($=G\MBH/\sigstar^2$) and $R_e$.

\subsection{Intrinsic Dispersion of the Correlations}
\begin{figure*}[t]
\centerline{ \epsfig{file=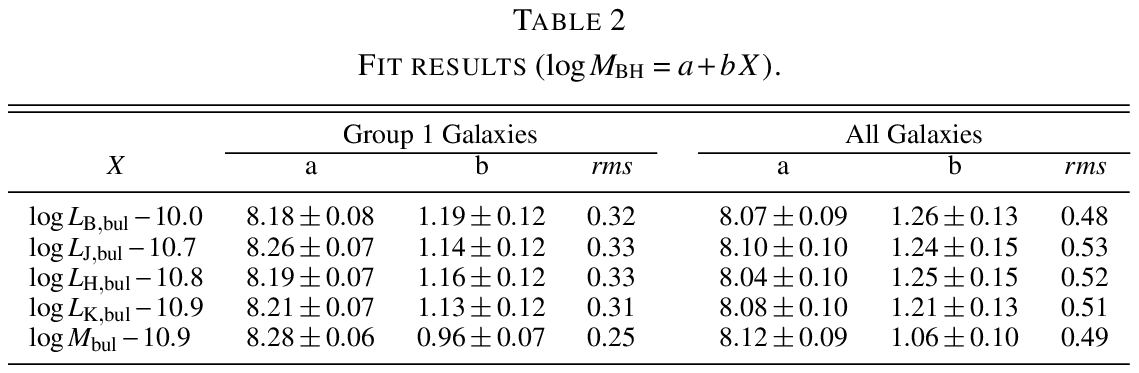} }
\end{figure*}

To compare the scatter of \MBH-\Lsph\ for different wavebands,
we have also analyzed the B-band bulge luminosities for our sample.
The upper limit of the intrinsic dispersion of the \MBH-\Lsph\ correlations
goes from $\sim 0.5$ dex in $\log\MBH$ when considering all galaxies, to $\sim
0.3$ when considering only those of Group 1.  Hence, for galaxies with
reliable \MBH\ and \Lsph,
the scatter of \MBH-\Lsph\ correlations is $\sim
0.3$, {\it independently of the spectral band used (B or JHK)}, comparable to that
of \MBH-\sigstar.
This scatter would be smaller if measurement errors are underestimated.
\cite{mclure02} and \cite{erwin} reached a similar conclusion 
using R-band \Lsph\ but on smaller samples. 
The correlation between R-band bulge light concentration and \MBH\
has a comparable scatter \citep{graham01}.

Since \MBH-\LBsph\ and \MBH-\LNIRsph\ have comparable dispersions, the 
rough bulge/disk decomposition (\S\ref{sec:intro}), the larger
reddening and stellar population effects do not apparently compromise the correlation.
Most of the galaxies in the sample are early types, and thus may be less
sensitive to the above effects.  However, the scatter in the \MBH-\Lsph\
correlations does not decrease significantly when considering only elliptical
galaxies.

The correlation between \MBH\ and \Msph\ has a slightly lower dispersion (0.25
versus 0.3) than \MBH-\Lsph.
If the scatter of \Msph-\Mdyn\ (0.21 dex) is an indication of the additional uncertainties on our virial
estimates, then the intrinsic scatter of \MBH-\Msph\ drops to $\sim 0.15$ dex.
Judging from the present data where only secure \MBH\ and \Lsph\ are included, 
\sigstar, \Msph, \LBsph\ and \LNIRsph\ provide equally 
good \MBH\ estimates to within a factor of $\sim 2$.

\subsection{Correlation slopes}

All the slopes are roughly unity, but those of
\MBH-\Lsph\ are systematically larger than that of \MBH-\Msph.  This is 
expected if \Lsph\ correlates with \MBH\ because of its dependence on \Msph\ 
through the stellar $M/L$ ratio. 
From our \Msph-\Lsph\ relation, we find that
$\log\mlr_K\,=\,0.18\log\LKsph-2.1$;
the weak dependence of $\mlr$ on $L$ 
fully accounts for the different slopes of \MBH-\LKsph\ and \MBH-\Msph, and
the same applies to the J and H bands.

All correlations are thus consistent with a direct proportionality between
\MBH\ and bulge mass. This contrasts with
previous claims of a non-linearity of the \MBH-\Msph\ relation \citep{laor01}
but is in agreement with \cite{mclure02}.
A partial correlation analysis of log\MBH\ (variable $x_1$),
log\sigstar\ ($x_2$), and log$R_e$ ($x_3$) shows that \MBH\ is {\it separately}
significantly
correlated both with \sigstar\ and $R_e$. The Pearson 
partial correlation coefficients, in which the known 
dependence of \sigstar\ and $R_e$
is eliminated, are $r_{12}=0.83$, $r_{13}=0.65$,
with a significance of $>99.9\%$.
This is shown graphically in Fig.\ \ref{fig:corr}c, where the
residuals of the T02 \MBH-\sigstar\ correlation are plotted against $R_e$;
there is a weak, but significant,  correlation of these residuals with $R_e$.
Consequently, when galaxy structural parameters 
are measured carefully from
2D image analysis, the additional, weaker, dependence of \MBH\ on $R_e$ is uncovered.
Thus, a combination of both \sigstar\ and $R_e$ is necessary to drive the
correlations between \MBH\ and other bulge properties.
This {\it fundamental plane} of black holes will be further investigated
elsewhere.

The average $\log \MBH/\Msph$ can be estimated assuming a log-normal
distribution with normally distributed observational errors.
With maximum likelihood
we find $\langle \log\MBH/\Msph \rangle = -2.63$
with an intrinsic dispersion of 0.27 dex (-2.79 and 0.49 dex 
for all galaxies). Adopting the method of \cite{mf01},
we find $\langle \log\MBH/\Msph \rangle = -2.81$ and $rms=0.36$ (-2.86 and 0.44 for all galaxies)
consistently with their result of -2.9 and 0.45 dex (see also \citealt{mclure02}).

\acknowledgments
We thank A.\ Capetti, E.\ Emsellem, W.\ Maciejewski, C.\ Norman, and E.\ Oliva  for useful discussions. This paper
was partially supported by ASI (I/R/112/01 and I/R/048/02) and
MIUR (Cofin01-02-02).  This publication uses the
LEDA database, the NASA/IPAC
Extragalactic Database operated by JPL, CalTech, under contract with
NASA, and data products from 2MASS, a
joint project of the University of Massachusetts and the IPAC/CalTech, funded
by NASA and NSF.

\end{document}